\documentclass[aps,prb,twocolumn,groupedaddress,floatfix,showpacs]{revtex4-2}
\usepackage{graphicx}
\usepackage{amsmath}
\usepackage{subcaption}
\usepackage{color}

\definecolor{green}{rgb}{0,0.5,0}
\begin{document}

%\preprint{}

\title{Quantum interference of Rydberg excitons in Cu$_2$0: quantum beats.}


%\thanks{}
\author{Sylwia Zieli\'{n}ska-Raczy\'{n}ska}
\author{David Ziemkiewicz}
\email{david.ziemkiewicz@utp.edu.pl}
 \affiliation{Department of
 Physics, Technical University of Bydgoszcz,
\\ Al. Prof. S. Kaliskiego 7, 85-789 Bydgoszcz, Poland}


\date{\today}


%\keywords{}


\begin{abstract}
A density matrix formalism is employed to calculate the emission of multi-level Rydberg excitonic system,  highlighting picosecond-scale dynamics and coherent effects such as quantum beats. The theoretical results for one and two-photon excitations of various Rydberg excitons and configurations providing the insight into their dynamics are compared with recent experimental data. In particular, the effect of Rydberg blockade on the quantum beat phenomenon is discussed.
\end{abstract}


%\keywords{}

\maketitle

\section{Introduction}

Bound states of electron-hole pairs in semiconductors demonstrate a hydrogen-like behavior in their high-lying excited states that are  known as Rydberg excitons (REs). The first observation of those quasiparticles  in Cu$_2$O in 2014  has opened a branch of semiconductor Rydberg physics, which however differs from their atomic analogue due to specific solid state environment; excitons are embedded in the background of a crystal lattice, which interact via screening and scattering.
\\
Rydberg excitons  have intensified usual excitonic properties including
dipole-dipole interactions that scale as $n^4$ and radiative lifetimes that scale as $n^3$ exceeding nanoseconds. For highly-excited states  with principal quantum number $n>>$1 the average electron-hole separation can reach $\mu$m. REs demonstrate strong Kerr type nonlinearieties for both visible light  (\cite{my2019},\cite{PRL2022})  and for microwaves  (\cite {Gallagher2022}, \cite{Pritchett2024}). A significant
effect concerning Rydberg excitons is Rydberg blockade, which is
a result of a long-range dipole-dipole and van der Waals
interactions between them.  These interactions can be strong enough to shift
the excitons energy  levels,  which prevents an exciton creation  in the
 vicinity of an already existing exciton.  So far the static spectroscopic properties of REs have been  examined in bulk and in nanoscale systems \cite{Hamid} proving their ability to be viable candidates for quantum technology. The  plurality of REs states in copper oxide offers the wealth of available quantum  (one or two-photon) excitations in the  wide range of frequencies from microwaves to optical ones and gives rise to  a huge amount of dynamical processes which can be studied in REs media.
In view of extraordinary properties of the REs  it seems appropriate  to look at their dynamics   due to their time-dependent interaction  with light.    
\\
In general, the light-matter interaction leads to interesting and relevant coherent and incoherent phenomena which are fundamental in the studies the properties of materials.  Investigations based on such interaction have revealed a plethora of information concerning excited quantum states, their coupling with light and dynamical time scales associated with fundamental processes in materials (for Cu$_2$O see \cite{Taylor,Naka17,Rotteger,Panda}).

One of coherent effects caused by such an interaction is the coherent excitonic beat, which is  an important spectroscopic signature providing information about excited quantum states. The phenomenon of quantum beating  relies on the concept of quantum mechanical superposition of  closely-lying excited states coupled by a near-resonant  laser pulse  with a ground state \cite{Scully}. The coherence of two or more excited states that is created by a short laser pulse manifests itself by modulation of the emitted radiation intensity.

Quantum beats were observed for the first time in 1955 \cite{Forrester}. Recently, experiments with quantum beats can be performed even on single photons  demonstrating
tunable Hong-Ou-Mandel interference in which the quantum beats of a single photon is observed \cite{Legerro}.

The observation of quantum beats yields a very direct information about the coherence properties of electronic or vibronic excitations, as was shown for atoms and molecules in many experiments \cite{Bitto,Han2021} as well for 1$S$ excitonic state \cite{Uihlein} and quadrupole polaritons in Cu$_2$O \cite{Frohlich1991}. They have become an important scientific tool for observing the time evolution of closely spaced energy levels \cite{Carter2000}. This makes them particularly well suited for the study of Rydberg excitons.

Grunwald \textit{et al} \cite{Grunwald} were the first who have analyzed the quantum coherent properties of Rydberg excitons absorption spectrum in Cu$_2$O and noticed the role of coherences between two adjacent upper excitonc states. Now Rydberg excitons explorations in Cu$_2$O have  reached the stage at which coherent quantum effects are observed (\cite{Thomas},\cite{Hamid2022}). This allows for a  controlled  quantum manipulation and state generation for excitons and opens  the possibility for developing precise and controllable optical devices on Rydberg excitons. As it was mentioned, Cu$_2$O provides the flexibility to operate at various frequencies creating Rydberg excitons and the separation between Rydberg levels is not much greater then the width of these levels therefore it seems to be a suitable medium for realizing quantum beats.  This phenomenon reflects an intrinsic quantum mechanical interference and can be applied to Rydberg excitonic physics providing detailed insight into the nature of excited states coupling. It should be mentioned that a quantum beat for Rydberg excitons of size reaching hundreds nanometers in Cu$_2$O, is a pure quantum effect, which  can be observed for objects of a mesoscopic scale.

In this paper we present the results of the time-dependent behavior of  Rydberg excitons excited by a short laser pulse leading to creation of quantum beats. Our calculations based on the master equation  \cite{ME1981}, in which quantum coherences are represented by off-diagonal elements of the density matrix. In Sec. II we set up the formalism for coherent exciton-state generation by  the pump pulse and emission of quantum beats. Next, in Sec. III we apply presented equations to calculate emission spectra exhibiting quantum beats in different cases of one- and two-photon excitations of Rydberg excitons in Cu$_2$O. Sec. IV is devoted to the analysis the ifluence of Rydberg blockade  on quantum beats spectra. Finally, we summarize our conclusions in Sec.V. Three Appendices contain calculation of dipole/quadrupole matrix elements, an estimation of exciton density and fits to the experimental energies and relaxation rates of $S,D$ exciton states.

\section{Theory}
Excitons are hydrogen-like excitations and in spite of the dependence of exciton energy on wavevector they may be considered as two-level systems (with a multiple excited state due to degeneracy). This is because in the absorption process the energy and wavevector are conserved so the resonant excitation with light will create the exciton in a state with well-defined energy and K. Here we focus our interest on the dynamical aspects and properties of Rydberg excitons in Cu$_2$O. We will  show that due to interaction with a  laser pulse  Rydberg excitonic states can be used to create the quantum  beats due to a superposition  between at least two different energy states. Quantum beats are manifested as damped oscillations  of radiation emitted by the system. These oscillations are the result of emission from a superposition of excited states created by a nearly resonant excitation.

\begin{figure}[ht!]
\centering
\includegraphics[width=.5\linewidth]{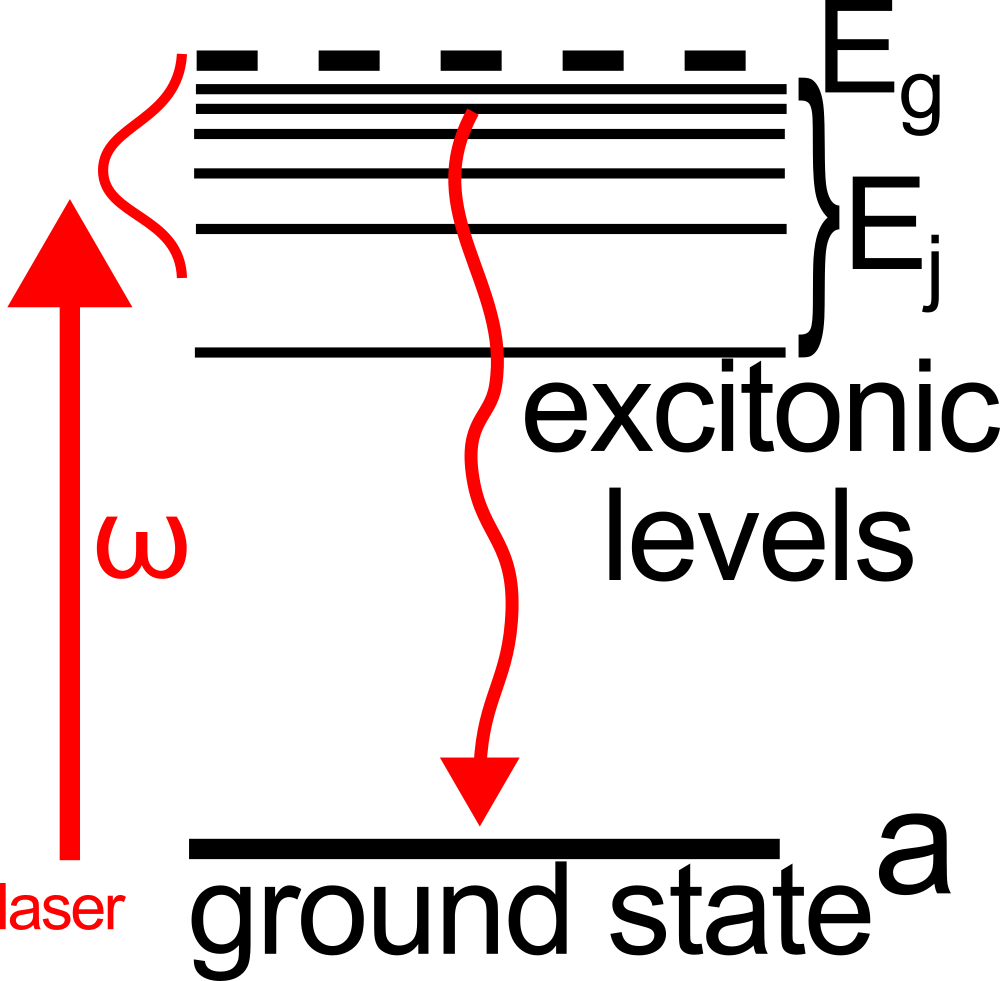}
\caption{General schematic of the system.}\label{fig:1}
\end{figure}

We consider a system  presented on Fig. 1, it consists of valence band as a ground state which is coupled  to at least two (or more) upper Rydberg excitonic states by a non-resonant laser pulse.
A single exciton resonance is modeled by two states, the excited state $|j\rangle$  and the excitonic vacuum as a ground state. Both states are coupled by the laser
 pulse of frequency $\omega$, amplitude $\varepsilon$ and polarization $\textbf{e}$
\begin{equation}\label{eq:field}
\textbf{E}(z, t)= \textbf{e}\varepsilon(z,t)\exp[i(\omega t-kz)].
\end{equation} 
The pulse is short and thus spectrally wide so it can couple several upper states at the same time as it is shown on Fig. \ref{fig:1}.
The time evolution of the excitonic density matrix $\rho=\rho(z,t)$ for an exciton at position $z$ is described by von Neumann equation with a phenomenological dissipative term R \cite{kossak, ME1981}
\begin{equation}\label{eq:vN}
i\hbar \dot{\rho}=[H,\rho]+R\rho.
\end{equation}
The Hamiltonian of such a system interacting with an electromagnetic wave  has the form 
\begin{eqnarray}
&&H=H_0+V+H_{vdW}=E_a|a\rangle \langle a|+\sum_j E_j|j \rangle \langle j|\nonumber\\
&&+\sum_j 2\Omega_j\hbar cos(\omega t)[|a\rangle \langle j|+|j \rangle \langle a|]+H_{vdW},
\end{eqnarray}
where 
\begin{equation}
\Omega_j=\frac{\varepsilon d_{aj}}{2\hbar}
\end{equation}
is the Rabi frequency for the $a \rightarrow j$ transition, $d_{aj}$ are transition dipole moments and $E_{j}$ are energies of exciton states. 
In Rydberg systems the dipole-dipole interaction  between excitons is responsible for a reduction of absorption, because of energy levels' shifts. The resulting effect of so-called  Rydberg blockade \cite{Kazimierczuk} occures,
 where a creation of another exciton is prevented in the vicinity of  the first exciton and an absorption bleaching is observed. The influence of Rydberg blockade results in strongly correlated many-body states of Rydberg excitations \cite{nature2021}, and is described by the van der Waals potential in the form
\begin{equation}
H_{vdW}=\sum\limits_{\mu}\frac{C^\mu_6}{R^6_{ij}}|\mu \rangle\langle\mu|,
\end{equation}
where $|\mu\rangle$ is a two-exciton eigenstate \cite{Walther2018}. It should be mentioned that particularly for excitonic states with a high principal quantum number $n$ the influence of the blockade is considerable since the coefficient $C_6 \sim n^{11}$, so that the van der Waals interaction leads to a significant shift of exciton energy levels. However, the number of created excitons depends on the laser intensity, therefore for lower exciton states created in laser fields of  low  intensities the blockade effect can be negligible, while for high excitonic states Rydberg blockade may be significant even for a very low laser power.
 
 In Cu$_2$O crystal the uppermost valence band has $\Gamma^+_5$ symmetry,  which is split into a lower band $\Gamma^+_8$ and an upper band $\Gamma^+_7$ by the spin-orbit interaction, and together with the lowest conduction bands $\Gamma^+_6$ and $\Gamma^+_8$ results in four exciton series (yellow, green, blue and violet). Due to symmetry properties of valence and conduction bands two types of excitation are possible - dipole allowed ($P$ excitons) and dipole forbidden, but quadrupole allowed ($S$ or $D$ excitons) because of the positive parity of both  the conduction  and  the valence bands. For dipole allowed (one-photon) transitions, the coupling elements are
\begin{eqnarray}
&&V_{aj}=-\varepsilon d_{aj},
\end{eqnarray}
$d_{aj}$ being dipole moment, while a dipole-forbidden excitation can be realized by two-photon transition through a virtual non-resonant state $v$ 
\begin{equation}\label{eq:vab}
V_{aj}=\frac{W_{av}W_{vj}}{E_a-E_v-\tilde{\omega}}, \qquad \tilde{\omega}=\frac{\omega}{2},
\end{equation}
where  $W_{av}=-\varepsilon d_{av}$. 
This effective second-order coupling given by Eq. (\ref{eq:vab}) follows from an adiabatic elimination of terms including the state $v$ from a complete set of equations which results from Eq. (\ref{eq:vN}) for a case of four-level system (which is an analog of Eqs (\ref{eq_glowne}), see below, completed  with the virtual state $v$ and additional couplings). If more non-resonant states $v_i$ give  contributions to the effective coupling a sum over $v_i$ in Eq.(\ref{eq:vab}) should be taken.
 
To illustrate and specify a theoretical approach we restrict our model to three-level system: valence band  \textit{a} and two  upper excitonic states \textit{b} and \textit{c}.
After making the rotating-wave approximation, i.e., transforming-off the terms oscillating rapidly  with optical frequency:  $\rho_{aj}=\sigma_{aj} e^{i\omega t}, \rho_{aa}=\sigma_{aa}$ $\rho_{jj}=\sigma_{jj}, j=b,c$ and  adding  terms describing relaxations within the system, von Neumann equation (\ref{eq:vN}) can be written in the form
\begin{eqnarray}\label{eq_glowne}
i\dot{\sigma}_{aa}&=&\Omega_{b}\sigma_{ba}+\Omega_{c}\sigma_{ca}-\Omega^*_{b}\sigma_{ab}-\Omega^*_{c}\sigma_{ac}\nonumber\\&+&i\Gamma_{ba}\sigma_{bb}+i\Gamma_{ca}\sigma_{cc}\nonumber\\
i\dot{\sigma}_{bb}&=&\Omega^*_{b}\sigma_{ab}-\Omega_{b}\sigma_{ba}-i\Gamma_{ba}\sigma_{bb}+i\Gamma_{cb}\sigma_{cc}\nonumber\\
i\dot{\sigma}_{cc}&=&\Omega^*_{c}\sigma_{ac}
-i\Gamma_{ca}\sigma_{cc}-i\Gamma_{cb}\sigma_{cc}\nonumber\\
i\dot{\sigma}_{ab}&=&\frac{1}{\hbar}(E_a+\hbar\omega-E_b)\sigma_{ab}+\Omega_{b}\sigma_{bb}+\Omega_{c}\sigma_{cb}\nonumber\\&-&\Omega_{b}\sigma_{aa}-i\gamma_{ab}\sigma_{ab}\nonumber\\
i\dot{\sigma}_{ac}&=&\frac{1}{\hbar}(E_a+\hbar\omega-E_c)\sigma_{ac}+\Omega_{c}\sigma_{cc}+\Omega_{b}\sigma_{bc}\nonumber\\&-&\Omega_{c}\sigma_{aa}-i\gamma_{ac}\sigma_{ac}\nonumber\\
i\dot{\sigma}_{bc}&=&\frac{1}{\hbar}(E_b-E_c)\sigma_{bc}+\Omega^*_{b}\sigma_{ac}-\Omega_{c}\sigma_{ba}-i\gamma_{bc}\sigma_{bc}.
\end{eqnarray}

As mentioned above, for a sufficient pump power, the excitonic energy levels $E_{b,c}$ will be subjected to an energy shift caused by the interaction $H_{VdW}$. Since the number of excitons in the system is a dynamic quantity, the energy levels will also shift in time, adding another degree of complexity to the system dynamics. The topic of Rydberg blockade is discussed in detail in  section IV.

The dissipative coefficients $\Gamma_{ij}, i \neq j$  describe a damping of exciton states and are determined by temperature-dependent homogeneous broadening due to phonons \cite{NJP} and broadening due to eventual impurities or structural imperfections. The relaxation damping rates for coherences are denoted by $\gamma_{ij}, i \neq j$; $\gamma_{ab}=\Gamma_{ba}/2 ,\gamma_{ac}=\Gamma_{ca}/2$ \cite{Artoni}. The coefficient $\gamma_{bc}$ is assumed to be negligible. For the particular values of $\Gamma_{ba}$, $\Gamma_{ca}$, we use the recent lifetime data from Ref. \cite{Thomas}.
It should be noted that only  relaxations inside the system are considered therefore the  total probability is conserved; $\sigma_{aa}+\sigma_{bb}+\sigma_{cc}=1$ and $\sigma_{aa}(0)=1$.

The time-dependent intensity of a light pulse emitted by an excited system is proportional to the mean  multipole (dipole $d_{ij}$ or quadrupole $q_{ij}$) moment $M$  between the ground and excited states,   $I(t)\sim|\langle M \rangle|^2 $, which is of the form
\begin{eqnarray}\label{exc_emis}
\langle M \rangle &=& Tr \rho M =(\sigma_{ab}M_{ba}+\sigma_{ac}M_{ca})e^{i\omega t}\nonumber\\
&+&(\sigma_{ba}M_{ab}+\sigma_{ca}M_{ac})e^{-i\omega t}\nonumber\\
&+&(\sigma_{bc}M_{cb}+\sigma_{cb}M_{bc}).
\end{eqnarray} 
It should be mentioned, that depending on whether the excitation is one- or two-photon due to selection rules the emitted pulse is an effect of a dipole or quadrupole transition. The estimation of elements for transition to the ground state is outlined in Appendix A.

Due to  of superposition of upper excited states,  contributions of which interfere, the emitted signal is modulated with the oscillation frequency
\begin{equation}
\omega_{cb}=\frac{2\pi}{T_{cb}}=\frac{E_c-E_b}{\hbar}.
\end{equation}

\section{Quantum beats with Rydberg excitons}
In the following subsections we will calculate the intensity of the emitted signal solving numerically Eqs (\ref{eq_glowne}-\ref{exc_emis}). The numerical results of above presented calculation scheme illustrate the response of Rydberg excitons in Cu$_2$O to a picosecond laser pulse that simultaneously excites a number of Rydberg states. In Eq. (\ref{eq_glowne}) there are several  parameters that need to be fitted to the properties of a particular excitonic system. First, one has the relaxation factors of the two types: $\Gamma_{ca}$, $\Gamma_{ba}$ corresponding to the transition to the ground state and the $\Gamma_{cb}$ corresponding to the transition between two excitonic levels. For $P$ excitons  the values of $\Gamma_{ja}$, $j=b,c...$ are adapted from the linewidth data presented in Ref. \cite{Kazimierczuk}; for $S$ and $D$ excitons dissipation rates are based on the recent experimental measurement of the exciton lifetimes \cite{Thomas} (see Appendix C). The relatively weak inter-excitonic transition rate $\Gamma_{cb}$ is calculated directly from the quadrupole transition moment, as outlined in Appendix A.
The couplings $V_{ij}$  depend on the electric field and the transition dipole or quadrupole moment for the specific transition. Again, these quantities are calculated in Appendix A.

On the other hand, the experimentally measured photoluminescence intensity depends heavily on nonradiative processes \cite{Naka18}. Therefore, we only calculate normalized emission intensity and make no attempt to estimate its absolute value, which may depend on many experimental conditions. Therefore, an accurate calculation of single photon quadrupole transition elements $q_{ij}$ to the ground state is omitted in favor of literature data, see Appendix A.

In Cu$_2$O, a near-resonant coupling which excites an exciton to any upper state provides the flexibility to operate at various frequencies between arbitrary excitonic states. This enables one to analyze at least three cases discussed below. In all considered examples we use realistic pulse laser parameters with the time duration of $\tau=4$ ps \cite{Thomas} and power density that is sufficiently low that the effects of Rydberg blockade can be neglected; the conditions in which blockade is essential are discussed in section \ref{sec_block}. In each calculation, the central frequency and the spectral width of the laser are adjusted to excite the specified excitonic states. We assume that the spectral width and pulse duration of the laser can be adjusted separately, and always the parameters are well above the energy-time uncertainty relation $\Delta E\Delta t \gg \hbar/2$.

\subsection{One-photon excitation}
\begin{figure}[ht!]
\centering
\includegraphics[width=.8\linewidth]{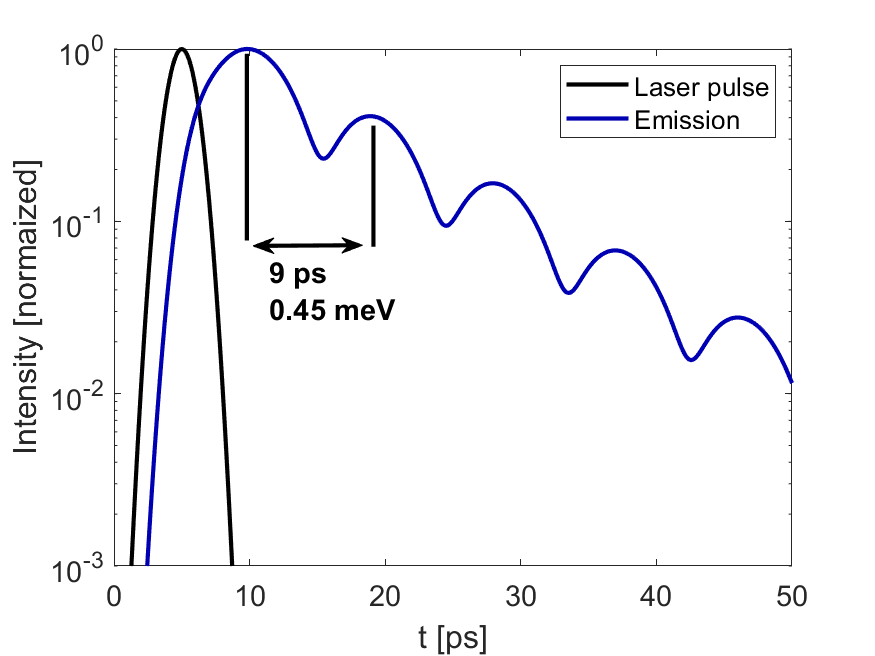}
\caption{Normalized emission intensity (in log scale) as a function of time for 7$P$-8$P$ system.}\label{fig:em1}
\end{figure}

As a first example, we consider a generic situation of a simple 3-level model  with a system depicted on Fig. \ref{fig:1}, where transition is dipole allowed.  The laser spectrum is adjusted so that only $7P$ and $8P$, which are very close to each other (with corresponding energies of $E_{7P}=2170.182$ meV and $E_{8P}=2170.635$ meV \cite{Kazimierczuk}), are excited and the quantum beats are realized between them.
 The pulse is covering both upper excitonic states and excites them simultaneously. The  result of the calculation is shown on Fig. \ref{fig:em1}. The logarithmic scale is used to show multiple beats clearly; in an experimental scenario only a few first oscillations, for $t<30$ ps, would be visible above the background noise.
The energy spacing between these states is thus $0.453$ meV which corresponds to the frequency $\omega \approx 0.7$ THz and the quantum beat oscillation period is $T \sim 9$ ps.
The emitted signal resembles a damped harmonic oscillator with the period $T_{ij}$, which agrees well with the energy splitting between exciton states. It it worth  stressing that such an oscillating behavior is an evident feature of the quantum coherences arising from coherent superposition of excitons. No quantum beats will be observed from any single excitons; they originate only from their coherent superposition.

As it was mentioned above, quantum beat investigations can be extended to a system with several closely spaced transitions, as it is in Rydberg excitons medium. One of the unique features of the recent pulsed laser experiments \cite{Thomas} is the fact that the pulses are relatively wide to cover a majority of the excitonic spectrum. This allows for a simultaneous observation of multiple excitonic lines and their complex quantum beat patterns.  To model such a situation one has to extend set of Eqs. (\ref{eq_glowne}) for more  excitons  and couplings, accordingly with Eq. (\ref{exc_emis}), which has to be expanded to allow for contributions from three $P$ excitons. As in the first case, the laser impulse has a duration of 4 ps but its spectrum is slightly wider to excite the 6$P$ exciton ($E_{6P}=2169.46$ meV) as well.

\begin{figure}[ht!]
\centering
\includegraphics[width=.9\linewidth]{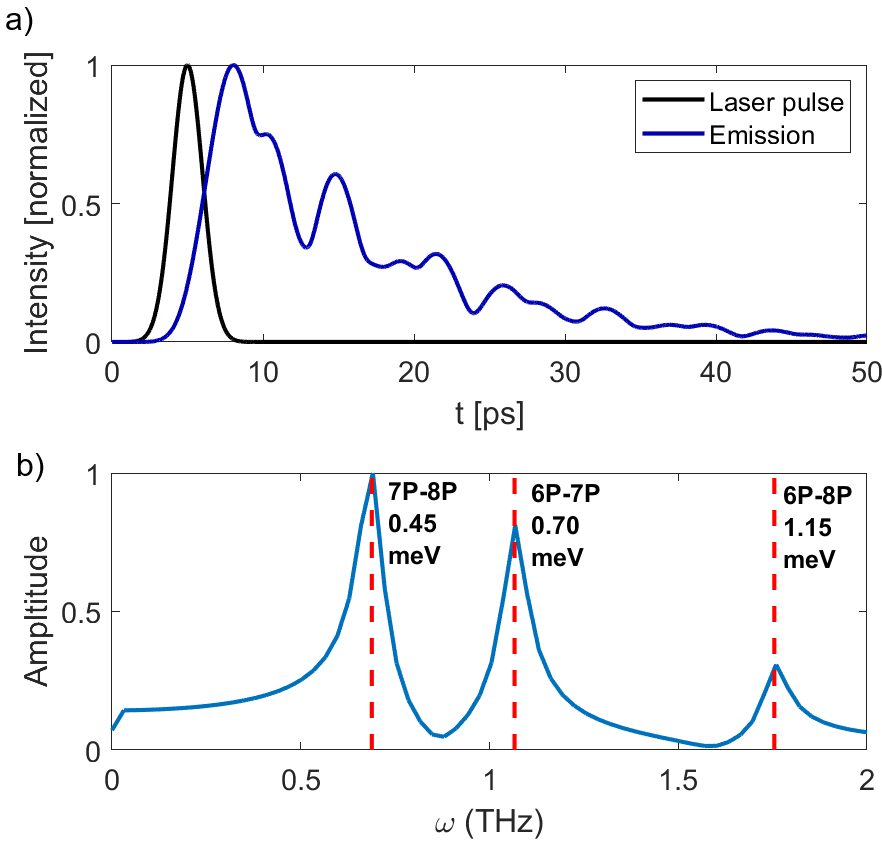}
\caption{a) Normalized emission intensity (linear scale) as a function of time for 6$P$-7$P$-8$P$ system. b) Frequency spectrum of the emission. Red, dashed lines indicate the individual beats corresponding to 6$P$-7$P$, 6$P$-8$P$, 7$P$-8$P$ energy level spacings.}\label{fig:em2}
\end{figure}

The results obtained for such a case are shown on Fig. \ref{fig:em2}a) and show a complicated pattern of multilevel quantum beats.
%\begin{figure}[ht!]
%\centering
%\includegraphics[width=.8\linewidth]{img/emis_multibeat}
%\caption{Emission of the $n=4$ S,D excitonic states, exhibiting complex quantum beat pattern.}\label{Fig:5}
%\end{figure} 
The amplitudes of individual beat frequency components depend on the relative transition dipole  moments of corresponding transitions. One can observe irregular oscillations which are outcome of interference from several beats originating from different superposition of excitonic states.\\ The frequency spectrum of the emitted light is shown on Fig.3 b). One can notice the frequencies matching the energy separation between nearby states. The dashed lines indicate the individual beats between relevant pairs of excitons.

\subsection{Quantum beats with two-photon excitation}
\begin{figure}[ht!]
\centering
\includegraphics[width=.5\linewidth]{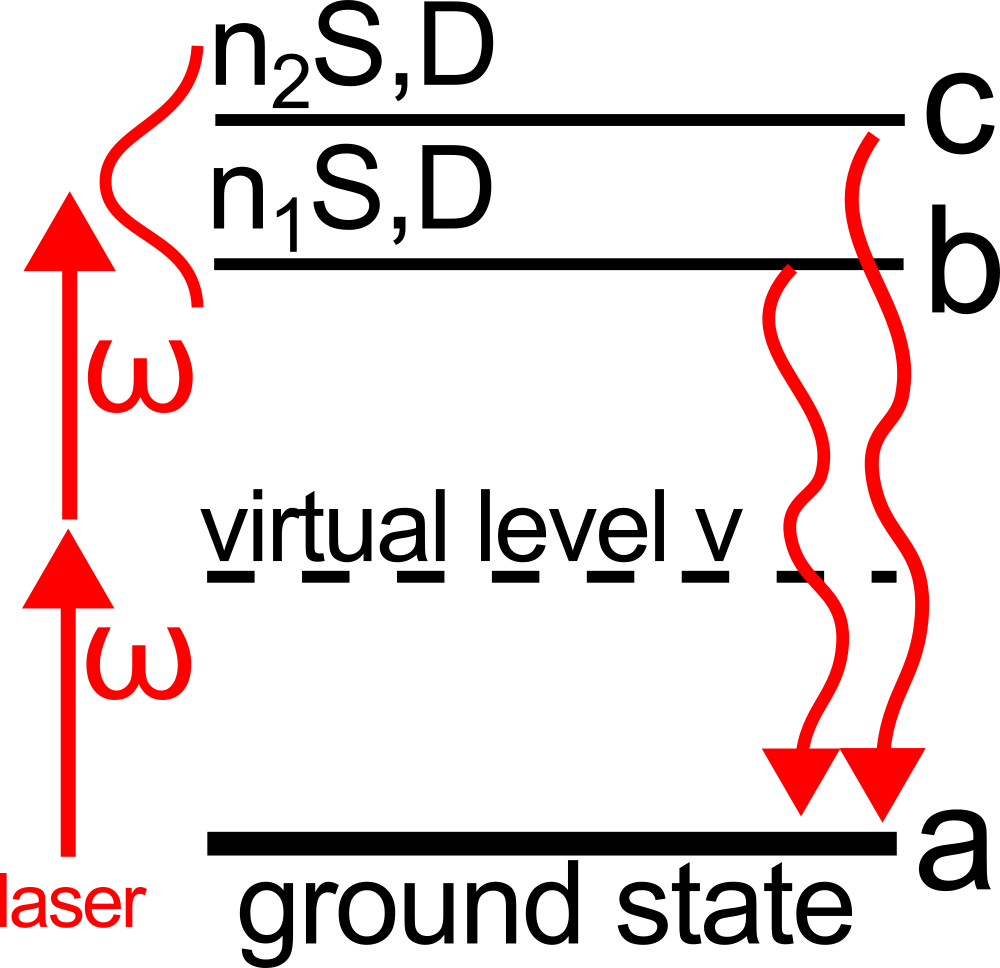}
\caption{Schematic of the system approximating the experiment from Ref. \cite{Thomas}.}\label{fig:1b}
\end{figure}

In this subsection, a two-photon excitation of even parity $S,D$ excitons is considered, with a subsequent emission through single photon quadrupole transition, as demonstrated in \cite{Thomas}. Specifically, a set of excitonic states is excited via non-resonant virtual level by two-photon absorption of infrared pulses and then the excitons recombine emitting visible light.
As a first example, one can consider two states $n_1$, $n_2$ with either $S$ or $D$ excitons (see Fig. \ref{fig:1b}).
Of particular relevance is the case of $n_1=n_2$, with upper $D$ state and lower $S$ state. Again, the laser pulse is sufficiently spectrally wide to couple both states at the same time.
The numerical calculations of Eqs. (\ref{eq_glowne}-\ref{eq:vab}) were performed for 
the energies of $S$ and $D$ exciton states  taken from \cite{Schweiner17}. Here, we consider power density for which the Rydberg blockade is weak enough that those energies can be assumed to be fixed; the inclusion of Rydberg blockade mechanism leading to dynamic shift of excitonic levels is discussed in  section IV.

\begin{figure}[ht!]
\centering
\includegraphics[width=.8\linewidth]{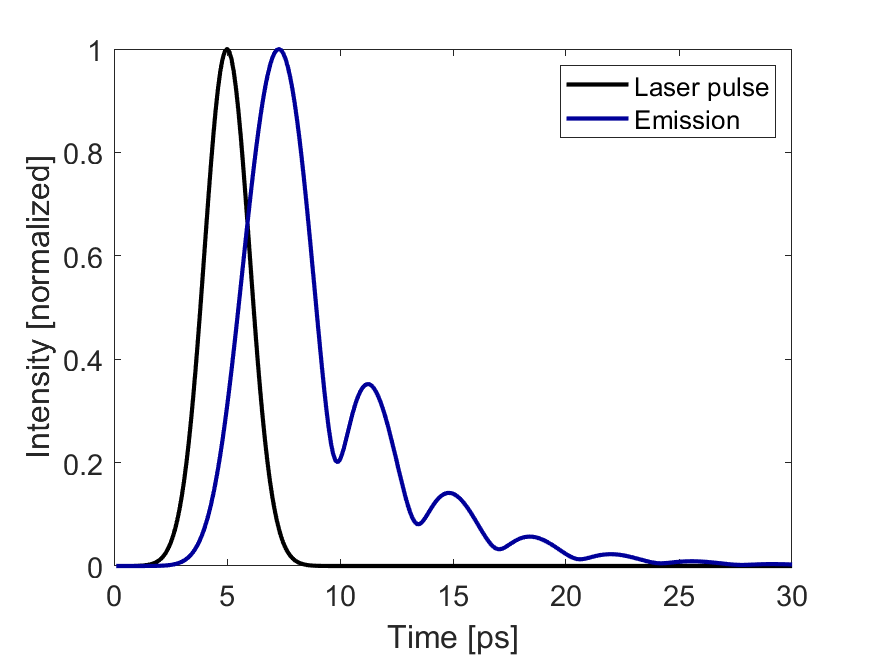}
\caption{Normalized emission intensity (linear scale) of n=5 $S, D$ excitonic states as a function of time.}\label{Fig:2}
\end{figure}

Fig. \ref{Fig:2} shows the calculated emission of the 5$S$-5$D$ exciton state pair. 
An incident laser pulse has a duration of $\tau_L \sim 4$ ps, similarly to the experiment in \cite{Thomas}. The absorption of the laser pulse results in an increase of exciton population, which then decays. However, the emission is not simply proportional to exciton population, but as can be seen from Eqs (\ref{eq_glowne}-\ref{exc_emis}), the emitted signal is a complex function of time-dependent coherences and dipole or quadrupole moments. As expected, the frequency of beats oscillations is in agreement with 5$S$-5$D$ levels separation, which is $E_{5D}-E_{5S}=0.56$ meV \cite{Schweiner17},   corresponding to $\omega=0.85$ THz and beats period of $7.3$ ps. The shape of emission, shown here in linear scale, is consistent with  the experimental results  \cite{Thomas}.

\begin{figure}[ht!]
\centering
\includegraphics[width=.8\linewidth]{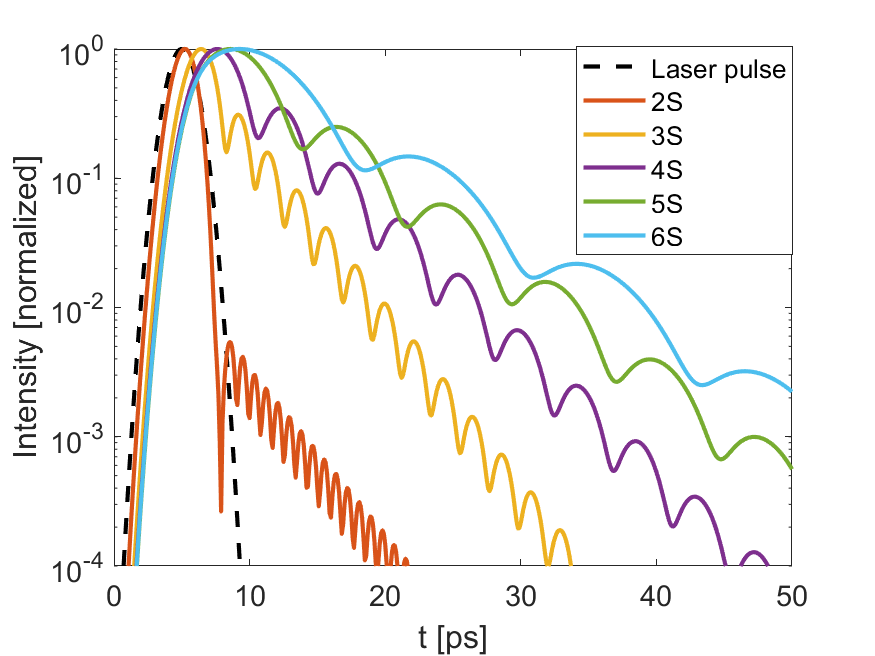}
\caption{Comparison of normalized emission intensity (log scale) of pairs of $S$,$D$ excitonic states with various principal quantum numbers.}\label{Fig:3}
\end{figure}

Similar calculation can be performed for various excitonic states. A comparison of emission intensities of various pairs of $S$,$D$ exciton states with different principal quantum numbers $n$ is shown on Fig. \ref{Fig:3}. As expected, the decay of the emission intensity is slower for higher $n$ due to the longer lifetimes. Moreover, because the energy spacing between $nS$ and $nD$ states decreases with $n$, the frequency $\omega_{ij}$ of quantum beats  correspondingly declines for highly excited states.

\subsection{Emission spectrum of multi-level system}

\begin{figure}[ht!]
\centering
\includegraphics[width=.8\linewidth]{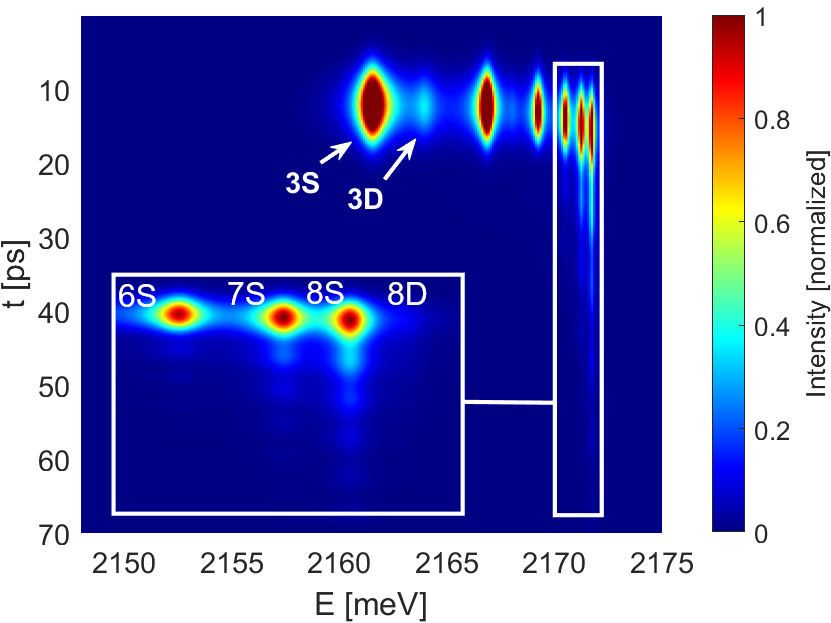}
\caption{Emission spectrum as a function of time and energy; inset: a zoom on the 6S-8S lines, highlighting quantum beats.}\label{Fig:6}
\end{figure}

The next step from calculating emission at a central frequency is to calculate the entire spectrum, so that one can follow the evolution of multiple emission lines of the system, similarly to the results presented in \cite{Thomas}. The normalized, central-frequency emission intensities calculated from Eq. (\ref{exc_emis}) are scaled according to the corresponding oscillator strengths \cite{Schweiner17}, describing emission intensity of individual excitonic states. A remarkable feature of Cu$_2$O is the fact that the $D$ exciton oscillator strength is particularly large, so that both $S$  and $D$ exciton lines are clearly visible in the emission spectrum \cite{Uihlein,Rogers2022}. We use the Lorentzian shapes of emission lines \cite{Rogers2022}, with linewidths according to recent measurements \cite{Thomas}. The complete emission spectrum, which is obtained from a sum of appropriately scaled emission lines, is then globally normalized. The laser is centered on the energy $E_0=2165$ meV and has the linewidth of 10 meV, which covers the majority of the exciton spectrum, from 3$S$ state to the band gap. The results are shown on Fig. \ref{Fig:6}.
The laser reaches the maximum intensity at $t=10$ ps. The emission intensity (color) reaches maximum slightly later with the delay increasing with $n$. As expected, states with larger principal quantum number are characterized by a longer emission time due to their longer lifetimes. In the case of $7S$ and $8S$ states, quantum beats (weak, secondary maxima) can be discerned at t=30-60 ps (Fig. \ref{Fig:6} inset). It is worth stressing that our theoretical calculations based on Eqs. (\ref{eq_glowne}-\ref{exc_emis}) with application of experimentally measured parameters for REs give the  similar emission spectra in the time-domain to these recently published by Chakrabarti \emph{et al} \cite{Thomas},  which validates our theoretical approach. Some differences between the calculation results and the measured spectra result from the lack of noise in calculations and  a rough estimation of dipole moments in our approach. 

\section{The effect of Rydberg blockade}\label{sec_block}
The pump laser power density and duration of the same order as in the recent experiments \cite{Thomas,Yoshioka}, result in exciton density on the order of $1/\mu$m$^3$ \cite{Yoshioka}. In such conditions, the blockade effects can become nontrivial, especially for highly excited states $n>10$. We also note that quantum beats are a well-known tool for precisely estimating small spacings between neighboring energy levels, which would be otherwise hard to measure \cite{Carter2000}. This is particularly relevant here since even $\mu$eV energy level shifts could be feasibly detected in experiments, greatly lowering the power necessary for the blockade effects to become visible.

For the case of $P$ excitons, the so-called blockade volume is given by \cite{Kazimierczuk}
\begin{equation}
V_{bl}=3 \cdot 10^{-7}n^7~\mu m^3.
\end{equation}  
For $n>10$, the blockade volume exceeds 3 $\mu$m$^3$ and the radius exceeds 1 $\mu$m, which means that at this distance, the energy shift caused by Rydberg blockade exceeds the linewidth of the particular state; for a monochromatic source of radiation  this means that further excitons cannot be efficiently created. However, in the case of a short, wideband pulse, the excitation is still possible. Therefore, instead of a reduction of absorption, one can possibly observe the shift of excitonic lines directly.

In general, dipole interactions between exciton with a quantum number $l>1$, such as $D$ excitons, are complex \cite{Assmann2020}, with individual exciton-exciton interaction energy that can be either positive or negative \cite{Singer2005}. This means that various excitonic states can experience redshift or blueshift that depends on the exciton density. For the case of Cu$_2$O, we follow the blockade potential calculations presented by Walther \textit{et al} \cite{Walther2018}. The interaction energy between two excitons with quantum number $n$, located at a distance $R_{ij}$ from each other is described by potential energy
\begin{equation}\label{eq_vi}
V_{ij}=\frac{C_6}{R_{ij}^6},
\end{equation}
with the constant $C_6 \sim n^{11}$ that is also dependent on the quantum numbers $l_1$, $l_2$ of both excitons \cite{Walther2018}. The total energy shift of a given exciton  is a sum over its interaction with all other excitons
\begin{equation}\label{eq_vi2}
V_i = \sum\limits_{j\neq i} \frac{C_6}{R_{ij}^6}.
\end{equation}
The total energy of a given exciton is thus $E_i' = E_i + V_i$, where, $E_i$ corresponds to $E_b$ or $E_c$ in Eqs. (\ref{eq_glowne}). In order to describe an entire ensemble of excitons, we calculate the average energy shift of the states $E_b$ and $E_c$. The procedure is as follows: we calculate the absorbed laser power and estimate the total number of excitons $N$, consistently with Ref. \cite{Yoshioka}, see Appendix B for details. Then, we divide the total population $N$ into populations $N_b$, $N_c$ proportionally to the density matrix elements $\sigma_{bb}$, $\sigma_{cc}$. Next step is Monte Carlo simulation \cite{Morawetz2023}, where the given numbers of excitons $N_b$, $N_c$ are distributed randomly in some specified volume. The energy $V_i$ is calculated from Eq. (\ref{eq_vi2}) for each exciton and the average is obtained for all excitons with energy level $E_b$ and $E_c$, leading to average shifted energies $E_b'$ and $E_c'$.

Apart from Monte Carlo simulation, one can make some simple analytical estimations; assuming that we have $N$ excitons that are distributed evenly in a given volume $V$, one can calculate the volume per exciton $V/N$ and a corresponding average radius of a sphere $\bar{R}$. Assuming that these spherical blockade volumes fill the available space using a perfect sphere packing, each sphere has 12 neighbors \cite{Wu2003}. This is the approximate number of excitons located within distance $\bar{R}$ of a given exciton. Since the blockade potential falls quickly with $\bar{R}$, only these nearest neighbors are relevant. Therefore, the total blockade energy can be estimated as
\begin{equation}\label{eq:sphere}
V_i \approx 12 \frac{C_6}{\bar{R}^6},
\end{equation}
where the exciton density  $\rho^{-1}=\frac{4}{3}\pi\bar{R}^3=\frac{V}{N}$.

For the 3-level systems considered here, one has two distinct excitonic states b,c and three possible interactions b-c, b-b, c-c between different excitons pairs. The total energy shift is calculated as a weighted average of relevant interaction energies, with weights proportional to level populations. For example, let's consider a system from Fig. \ref{fig:1b}, where levels b and c correspond to $S$ and $D$ exciton. To calculate the energy shift of the $S$ exciton, one has to consider $S-S$ and $S-D$ interaction and use the populations of $S$ and $D$ excitons to calculate the weighted average. Correspondingly, an energy shift of $D$ exciton depends on $S-D$ and $D-D$ interactions.

\begin{figure}[ht!]
\centering
\includegraphics[width=.8\linewidth]{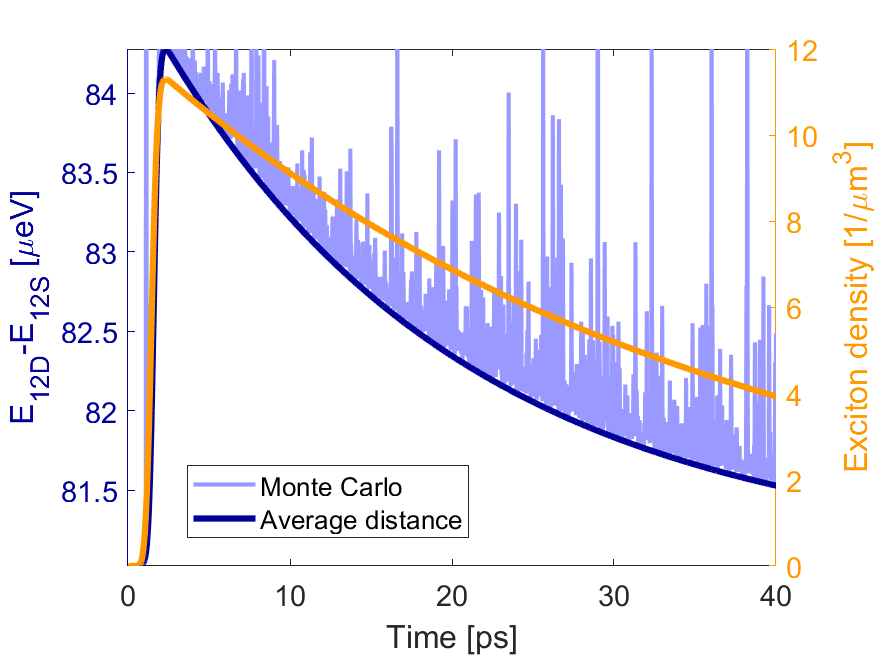}
\caption{Exciton density and energy spacing $E_{12D}-E_{12S}$ as a function of time.}\label{Fig:8}
\end{figure}

For illustration let's consider $12S$ and $12D$ exciton states, with energy of $E_{12S}\approx 2171.40$ meV and $E_{12D} \approx 2171.48$ meV \cite{Gallagher2022_b}. The power density is assumed to be of the same order as in \cite{Yoshioka}. The results are shown below on Fig. \ref{Fig:8}.
The exciton density reaches a maximum value of $\rho \sim 10/\mu$m$^3$, corresponding to exciton-exciton distance of $\sim 0.3$ $\mu$m. The energy $E_{12D}-E_{12S}$ increases by $3.5$ $\mu$eV, which is a $4\%$ change that could be detectable in quantum beat pattern. Specifically, as the exciton density decreases, the period of beat oscillations will increase from approximately 49 ps to 51 ps, allowing one to observe subtle, dynamic effects of Rydberg blockade.

 The blockade-induced energy shift has been calculated both with Eq. (\ref{eq:sphere}) assuming the sphere packing and with Monte Carlo simulation. Notably, Monte Carlo results predict a slightly higher energy due to the fact that roughly half of randomly distributed excitons are located closer to each other than the sphere packing predicts; due to the fast $1/R^6$ scaling, an increase of energy caused by smaller than average $R<\bar{R}$ is larger than decrease for $R>\bar{R}$. Due to the random nature of Monte Carlo method, the obtained energy contains a lot of noise with sharp peaks corresponding to exciton pairs located particularly close to each other. We note that the random spread of energy shifts reflects real experimental situation where excitons are distributed randomly; instead of a singular shift energy, there will be a range of most probable energy shifts, leading to a complicated quantum beats pattern.

\section{Conclusions}
We have presented the method based on evolution of density matrix to calculate the quantum beats in Rydberg excions medium. Coherent quantum beats are an important phenomenon particularly useful in understanding of  the coherent light-matter interaction and allowing the observation of coherent quantum effect in Rydberg excitons medium. Quantum beats, which manifest as oscillations in the emission intensity, enable one to observe the dynamics of highly excited state transitions in Cu$_2$O. We should stress that a quantum beat is a pure quantum phenomenon, which in case of Rydberg excitons, can be observed for objects of huge (in the quantum scale) dimensions, reaching hundreds of nanometers.
We have analyzed several cases of one or two-photon excitations of $P$, $S$ and $D$ Rydberg excitons, respectively for various excitation frequencies and pulse linewidths. The calculated emission spectra are in a good agreement with recent experiments \cite{Thomas}. Crucially, our approach is suitable for modeling quantum beats that are evident in measured spectra. These oscillations, with the period given by the inverse transition energy between the excited levels, allow one to precisely estimate the separation between the levels and observe how it evolves in time. This may be especially useful for possible future studies of Rydberg blockade; unlike traditional, indirect observation of this phenomenon through optical bleaching, quantum beats allow one to measure the blockade-induced energy shifts directly, even in the case where these shifts are much smaller than exciton linewidths.

\section{Appendix A: calculation of dipole and quadrupole moments}
Many of the key properties of the excitonic system can be estimated with first-principles calculations using hydrogenlike wavefunctions in the form
\begin{equation}
\Psi_{nlm}=R_{nm}(r)Y_{lm}(\theta,\phi)
\end{equation}
with
\begin{eqnarray}
&&R_{nl}=\sqrt{\left(\frac{2}{na^*}\right)^3\frac{(n-l-1)!}{2n(n+l)!}}e^{-r/na^*}\nonumber\\&&\left(\frac{2r}{na^*}\right)^lL_{n-l-1}^{2l+1}\left(\frac{2r}{na^*}\right)
\end{eqnarray}
with Bohr radius $a^* = 1.1$ nm and generalized Laguerre polynomial $L$. The angular part
\begin{equation}
Y_{l,m}(\theta,\phi)=(-1)^m \sqrt{\frac{2l+1}{4\pi}\frac{(l-m)!}{(l+m)!}}P_l^m(\cos(\theta))e^{im\phi}
\end{equation}
with associated Legendre polynomial $P_l^m$.

For the purpose of initial verification of the results, we will follow the experimental data of Kazimierczuk et al \cite{Kazimierczuk}. In particular, we assume the same Bohr radius $a^*=1.1$ nm and other relevant material parameters.

First, one can estimate the transition dipole moment of various inter-excitonic transitions, as well as the transition to the ground state. Specifically, the transition moment is
\begin{eqnarray}
d_{12}&=&\langle \Psi_{n,l,m}|er|\Psi_{n',l',m'}\rangle \nonumber\\ &=& \int\limits_{r=0}^\infty R_{nm}^*R_{n'm'}r^2 er dr \int\limits_{\theta=0}^\pi \int\limits_{\phi=0}^{2\pi} Y_{l,m}^*Y_{l',m'} \sin\theta d\theta d\phi\nonumber\\
\end{eqnarray}
where $\vec r$ is projected onto appropriate axis. For the low angular momentum states (S,P,D excitons), the angular part can be readily calculated analytically; for the radial part, especially in the context of high $n$ states, numerical calculation is used. As a first example, let's consider a radiative transition $P_z \rightarrow S$, $l=1$, $l'=0$, $m=m'=0$, $r_z=r\cos\theta$
\begin{eqnarray}
&&\int\limits_{\theta=0}^\pi \int\limits_{\phi=0}^{2\pi} Y_{10}^*Y_{00} \sin\theta\cos\theta d\theta d\phi = \frac{1}{\sqrt{3}} 
\end{eqnarray}
which is the result used in the work \cite{maser_OL}. The table 1 contains the transition dipole moments calculated with the above factor and numerically integrated radial function overlap. 

\begin{table}[ht!]
\centering
\begin{tabular}{|l|l|l|l|l|l|l|l|}
\hline
n1S\textbackslash{}n2P & 2    & 3     & 4     & 5     & 6     & 7     & 8     \\ \hline
1                      & 1.29 & 0.52  & 0.30  & 0.21  & 0.15  & 0.12  & 0.10  \\ \hline
2                      & 5.19 & 3.06  & 1.28  & 0.77  & 0.54  & 0.41  & 0.32  \\ \hline
3                      & 0.94 & 12.72 & 5.46  & 2.26  & 1.36  & 0.95  & 0.72  \\ \hline
4                      & 0.38 & 2.44  & 23.21 & 8.51  & 3.45  & 2.06  & 1.44  \\ \hline
5                      & 0.23 & 0.97  & 4.60  & 36.71 & 12.20 & 4.86  & 2.88  \\ \hline
6                      & 0.16 & 0.57  & 1.79  & 7.40  & 53.19 & 16.54 & 6.50  \\ \hline
7                      & 0.12 & 0.40  & 1.04  & 2.83  & 10.86 & 72.67 & 21.53 \\ \hline
8                      & 0.09 & 0.30  & 0.72  & 1.64  & 4.10  & 14.96 & 95.15 \\ \hline
\end{tabular}
\caption{Transition dipole moments for $n_1S \rightarrow n_2P$ transitions and various $n_1$, $n_2$, in multiples of $ea^*$.}
\end{table}

The results are consistent with \cite{maser_OE}. As expected, the dipole moment is roughly proportional to exciton size $\sim n^2$ and is the largest for $nS \rightarrow nP$ transition. 

Now, we can proceed to calculate the quadrupole transitions necessary for comparison with the recent results in \cite{Thomas}. The quadrupole transition moment is given by \cite{Jackson}
\begin{equation}
(q_{12})_{ij}=\langle \Psi_{n,l,m}|e(r_ir_j-\frac{1}{3}r^2\delta_{ij})|\Psi_{n',l',m'}\rangle
\end{equation}
Specifically, for $D \rightarrow S$ transition $D_{zz} \rightarrow S$, $l=2$, $l'=0$, $m=m'=0$, $r_z=r\cos\theta$. The only nonzero element is $q_{zz}=ez^2=er^2\cos^2\theta$. The results are summarized in table 2. 
\begin{table}[ht!]
\begin{tabular}{|l|l|l|l|l|l|l|}
\hline
n1S\textbackslash{}n2D & 3     & 4      & 5      & 6       & 7       & 8  \\ \hline
1                      & 0.63  & 0.41   & 0.29   & 0.22    & 0.17    & 0.14  \\ \hline
2                      & 18.99 & 2.85   & 0.93   & 0.41    & 0.22    & 0.13 \\ \hline
3                      & 51.34 & 71.13  & 14.77  & 6.22    & 3.49    & 2.28     \\ \hline
4                      & 16.64 & 193.60 & 184.81 & 41.15   & 18.10   & 10.46  \\ \hline
5                      & 5.07  & 64.14  & 506.18 & 394.78  & 89.60   & 39.79  \\ \hline
6                      & 2.69  & 18.35  & 169.90 & 1086.58 & 743.35  & 169.22  \\ \hline
7                      & 1.75  & 9.39   & 46.37  & 367.80  & 2053.87 & 1280.47  \\ \hline
8                      & 1.28  & 6.00   & 23.05  & 96.84   & 699.35  & 3548.75 \\ \hline
\end{tabular}
\caption{Transition quadrupole moments for $n_1S \rightarrow n_2D$ transitions and various $n_1$, $n_2$, in multiples of $e(a^*)^2$.}
\end{table}

A striking feature of the results is that while the quadrupole moments are smaller than dipole moments by a factor of $\sim a^*$, they increase with $n$ more quickly. Therefore, for sufficiently highly excited states, the dynamics of the system can be dominated by quadrupole transitions.

To calculate the transition rate $\Gamma_{bc}$ used in Eqs. (\ref{eq_glowne}), we use the Einstein A coefficient for spontaneous emission in quadrupole transition
\begin{equation}
A_{ij}^q = \frac{|q_{ij}|^2\omega_{ij}^5}{40 \pi \epsilon_0 \epsilon_b \hbar (c/\sqrt{\epsilon_b})^5}
\end{equation}
with $c/\sqrt{\epsilon_b}$ being the speed of light in Cu$_2$O. Then, we proceed to calculate the $B$ coefficient for stimulated emission
\begin{equation}
B_{ij}=\Gamma_{ij}=\frac{\pi^2c^3}{\hbar\omega_{ij}^3}A_{ij}.
\end{equation}

Finally, the quadrupole moments of the transition between the excitonic state and the ground state are necessary for the emission process. In this paper, we only use normalized emission and assume that its rate is very weak, so that it doesn't affect the dynamics of state populations since the dynamics of the system are dominated by nonradiative processes \cite{Naka18}. In particular, the oscillator strength of the transition to the $1S$ state is estimated as $10^{-9}$ \cite{Frohlich1991}, which is approximately 6 orders of magnitude less than for dipole-allowed $2P$ state. Exact transition rate can be obtained by considering quadrupole moment between relevant excitonic state and Bloch function of the valence band electron \cite{Heck017}. For an extensive  discussion of the limitations of hydrogenlike model, detailed \emph{ab initio} calculation of the transition rates to the ground state and  calculations of dipole and quadrupole oscillator strengths, see \cite{Schweiner2017}.
 
\section{Appendix B: estimation of exciton density}
In order to estimate the relation between input power and exciton density, the efficiency of two-photon excitation process needs to be known. For that, we refer to \cite{Yoshioka}, where a pulse with intensity of 6.1 MW/cm$^2$ was applied to excite the  dark exciton states in Cu$_2$O in the sample thickness of 800 $\mu$m. Assuming that the pulse is completely absorbed, the average power density is $\rho_P \approx 0.076$ mW/$\mu$m$^3$. The pulse duration is $\tau=2$ ps. The energy density absorbed by the medium is thus $\rho_E=\rho_P\tau \approx 950$ eV/$\mu$m$^3$. The exciton density estimated in \cite{Yoshioka} is $2.5 \cdot 10^{11} cm^{-3} = 0.25 \mu m^{-3}$. Therefore, the volume per single exciton is 4 $\mu$m$^3$, corresponding to a sphere radius of approximately 1 $\mu$m, which is the mean inter-exciton distance \cite{Yoshioka}.

\begin{figure}[ht!]
\centering
a)\includegraphics[width=.8\linewidth]{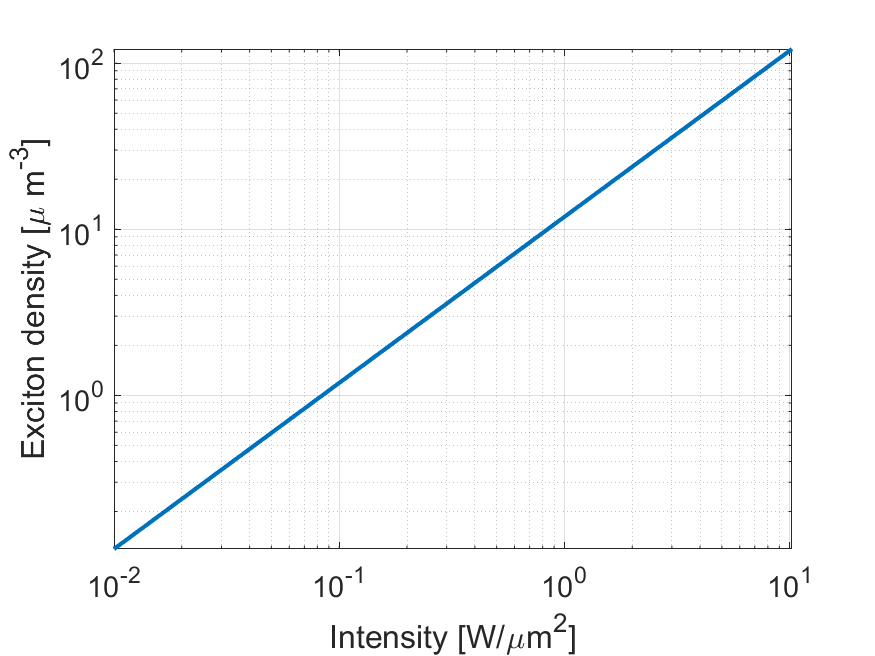}
b)\includegraphics[width=.8\linewidth]{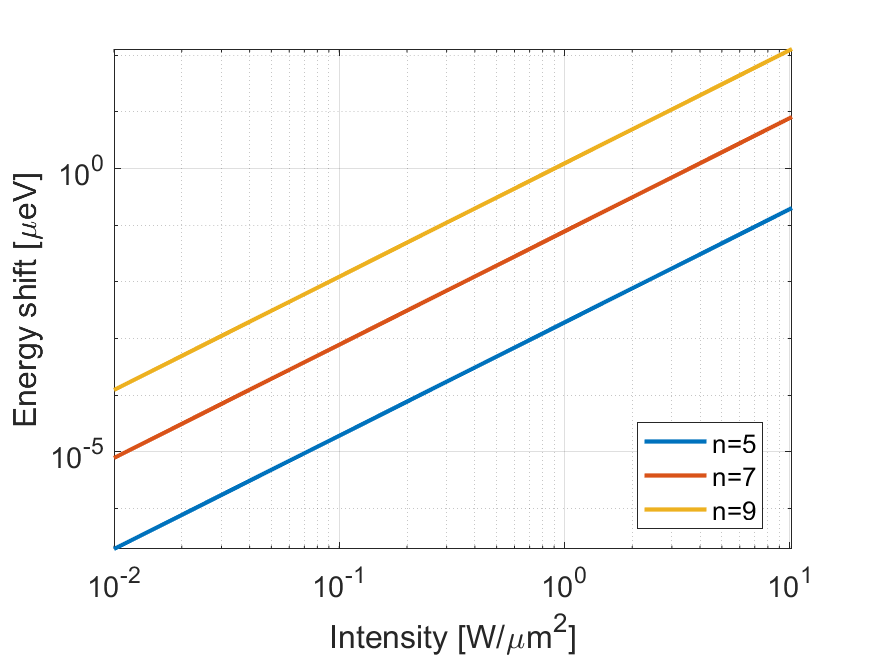}
\caption{a) Exciton density as a function of pulse intensity; b) Energy shift as a function of intensity, calculated for S excitons.}\label{fig_density}
\end{figure}

The energy necessary to create one exciton is $E \approx 2.1$ eV. Therefore, to obtain the exciton density of $0.25 \mu m^{-3}$, one needs energy density of $\rho_E' \approx 0.525$ eV/$\mu$m$^3$. By comparing $\rho_E$ and $\rho_E'$, we get an efficiency
\begin{equation}
\eta = \frac{0.525}{950}=5.5 \cdot 10^{-4}.
\end{equation} 
Now, we can calculate the exciton density as a function of pump power. It is given by
\begin{equation}
\rho_e = \frac{\eta}{V}\frac{\int\limits_{t=0}^\tau P(t)dt}{E_{exc}},
\end{equation}
where $V$ is the medium volume absorbing the pulse, $P(t)$ is the pump power, $\tau$ is pulse duration and $E_{exc}$ is exciton energy. In the case where exciton lifetime $t_{exc}$ is comparable to pulse duration, one has to use a convolution to get a modified function
\begin{equation}
P'(t)=\int\limits_{-\infty}^\infty P(\tau)e^{-(t-\tau)/t_{exc}}d\tau.
\end{equation}
Given some exciton density $\rho_e$, the volume per exciton is $V_e=\rho_e^{-1}$ and the radius of equivalent spherical volume is
\begin{equation}
R=\left(\frac{3V_e}{4\pi}\right)^{1/3}.
\end{equation}
Assuming the blockade potential $V_{VdW}=C_6/R^6$, one can write
\begin{equation}
V_{VdW}=\frac{C_6}{(\frac{3V_e}{4\pi})^2}=\frac{16}{9}\pi^2C_6\rho_e^2.
\end{equation}
Therefore, the energy shift due to Rydberg blockade is proportional to the square of exciton density, which in turn is a product of pump power and duration. The dependence of exciton density on laser intensity, calculated for sample thickness $L=800$ $\mu$m and pulse duration $\tau=4$ ps, is shown on Fig. \ref{fig_density} a). The Fig. \ref{fig_density} b) depicts the energy shift caused by Rydberg blockade, as a function of intensity. The strong dependence on quantum number $n$ is evident.
 
\section{Appendix C: Energy and lifetime fits}

\begin{figure}[ht!]
\centering
\includegraphics[width=.8\linewidth]{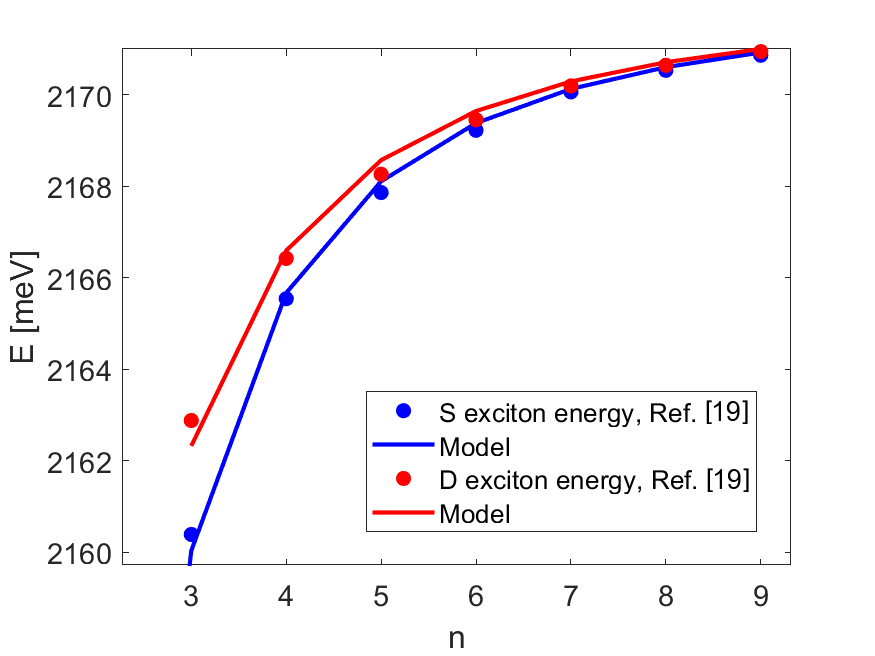}
\caption{The exciton energy as a function of principal quantum number n. Experimental data from \cite{Thomas} (dots) and fits (lines) are shown.}\label{Fig:AC1}
\end{figure}

Basing on the experimental data in \cite{Thomas} and \cite{Schweiner17}, we use an expression for the energy of S excitonic state
\begin{equation}
E_{n,S}=E_g-\frac{R^*}{(n-\delta_s)^2}
\end{equation}
with $R^*=87.8$ meV, $E_g=2172.08$ meV and quantum defect $\delta_s=0.3$; for D states, analogously
\begin{equation}
E_{n,D}=E_g-\frac{R^*}{(n-\delta_d)^2}
\end{equation}
with $\delta_d=0$. We use the dominant $\Gamma_5^+$ state that has significantly higher oscillator strength than other states \cite{Schweiner17}. The fitted quantum defect values are close to the estimations in \cite{Thomas} that are based on lifetime fits. The results are shown on Fig. \ref{Fig:AC1}.

\begin{figure}[ht!]
\includegraphics[width=.8\linewidth]{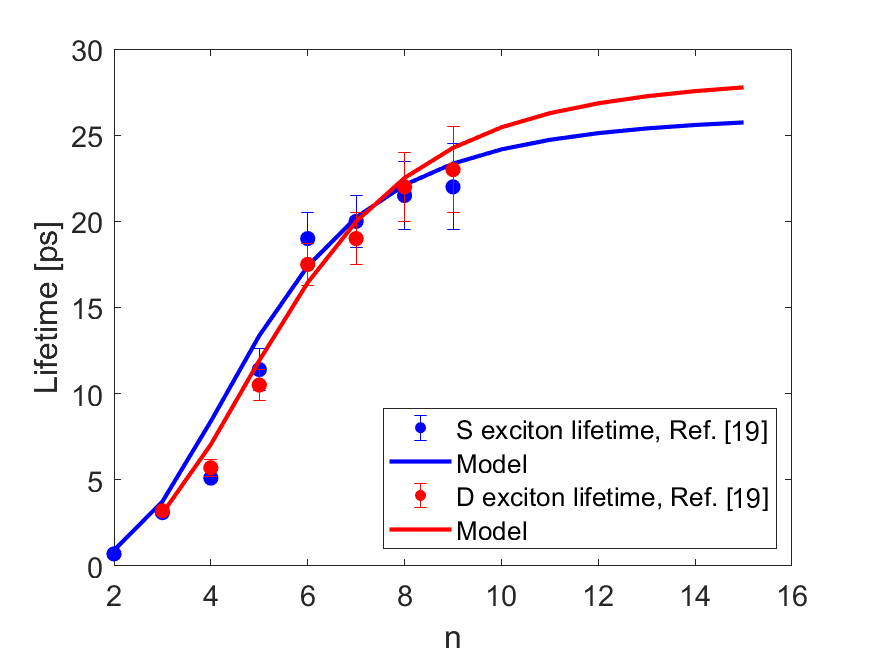}
\caption{The exciton lifetime as a function of principal quantum number n. Experimental data from \cite{Thomas} (dots) and fits (lines) are shown.}\label{Fig:AC2}
\end{figure}

The phenomenological damping constants representing the decay of exciton population are also fitted to the measurements in \cite{Thomas} (Fig. \ref{Fig:AC2}). Specifically, for S excitons
\begin{equation}
\Gamma_{n,S} = \frac{1}{\tau_{n,S}} = \frac{4}{(n-\delta_S)^{3.3}}+0.025~\mbox{[meV]},
\end{equation}
and for D excitons
\begin{equation}
\Gamma_{n,D} = \frac{1}{\tau_{n,D}} = \frac{9}{(n-\delta_D)^{3.5}}+0.023~\mbox{[meV]}.
\end{equation}
In the limit of large $n$, these values approach $0.025$ and $0.023$ meV, which corresponds to $\tau_{n,S} \approx 26.3$ ps and $\tau_{n,D} \approx 28.6$ ps.

\end{document}